\documentclass[sigconf]{acmart}

\copyrightyear{2024}
\acmYear{2024}
\setcopyright{rightsretained}
\acmConference[C\&C '24]{Creativity and Cognition}{June 23--26, 2024}{Chicago,
IL, USA}
\acmBooktitle{Creativity and Cognition (C\&C '24), June 23--26, 2024, Chicago,
IL, USA}
\acmDOI{10.1145/3635636.3656204}
\acmISBN{979-8-4007-0485-7/24/06}

\usepackage{colortbl}
\usepackage{graphicx}
\usepackage{enumitem} 

\newcommand{\killpunct}[1]{}
 
\begin{document}

\title[{Homogenization Effects of Large Language Models on Human Creative Ideation}]{Homogenization Effects of Large Language Models\\on Human Creative Ideation}

\author{Barrett R. Anderson}
\affiliation{
\institution{Independent Researcher}
\city{Santa Cruz}
\state{California}
\country{USA}
}
\email{barrettrees@gmail.com}

\author{Jash Hemant Shah}
\affiliation{
\institution{Santa Clara University}
\streetaddress{500 El Camino Real}
\city{Santa Clara}
\state{California}
\country{USA}
}
\email{jshah5@scu.edu}

\author{Max Kreminski}
\affiliation{
\institution{Santa Clara University}
\streetaddress{500 El Camino Real}
\city{Santa Clara}
\state{California}
\country{USA}
}
\email{mkreminski@scu.edu}


\begin{abstract}
Large language models (LLMs) are now being used in a wide variety of contexts, including as creativity support tools (CSTs) intended to help their users come up with new ideas. But do LLMs actually support user creativity? We hypothesized that the use of an LLM as a CST might make the LLM's users feel more creative, and even broaden the range of ideas suggested by each individual user, but also homogenize the ideas suggested by different users. We conducted a 36-participant comparative user study and found, in accordance with the homogenization hypothesis, that different users tended to produce less semantically distinct ideas with ChatGPT than with an alternative CST. Additionally, ChatGPT users generated a greater number of more detailed ideas, but felt less responsible for the ideas they generated. We discuss potential implications of these findings for users, designers, and developers of LLM-based CSTs.
\end{abstract}

\ccsdesc[500]{Human-centered computing~Empirical studies in HCI}
\ccsdesc[300]{Applied computing~Arts and humanities}
\ccsdesc[100]{Computing methodologies~Natural language processing}

\keywords{creativity support tools, divergent ideation, large language models, user study}

\begin{teaserfigure}
  \includegraphics[width=\textwidth]{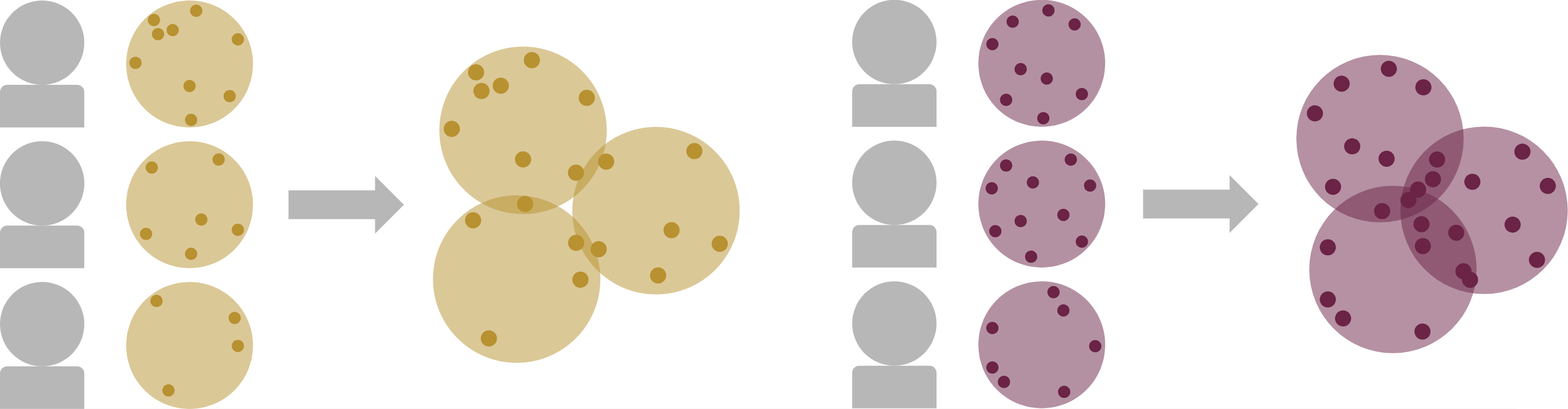}
  \caption{Homogenization analysis involves semantic similarity comparisons between artifacts produced by users of creativity support tools (CSTs). We apply homogenzation analysis to two different CSTs for divergent ideation, and find that users of the Oblique Strategies deck (on the left) and ChatGPT (on the right) each produce similarly homogenous sets of ideas as \emph{individuals}---but collectively, users of ChatGPT produce a more homogenous set of ideas at the \emph{group} level (as shown by the higher degree of overlap between the sets of ideas produced by each user).}
  \Description{Homogenization analysis involves semantic similarity comparisons between artifacts produced by users of creativity support tools (CSTs). We apply homogenzation analysis to two different CSTs for divergent ideation, and find that users of the Oblique Strategies deck (on the left) and ChatGPT (on the right) each produce similarly homogenous sets of ideas as \emph{individuals}---but collectively, users of ChatGPT produce a more homogenous set of ideas at the \emph{group} level (as shown by the higher degree of overlap between the sets of ideas produced by each user).}
  \label{fig:HomogenizationAnalysis}
\end{teaserfigure}

\maketitle

\section{Introduction}
Success in creative contexts (such as creative writing and product design) often hinges on the ability to come up with ideas that are---in line with prominent creativity researcher Margaret Boden's definition of creativity---simultaneously ``new, surprising, and valuable'' to some extent~\cite{Boden}. Though research in \emph{creativity support tools} (CSTs) has long aimed to produce software systems that can support parts of human creative processes~\cite{ShneidermanCSTs,CSTReviewFrich,CSTReviewChung}, it was largely not until the recent wave of developments in large language models (LLMs) that software tools capable of directly generating potentially creative ideas in arbitrary domains began to seem feasible. Especially since the release of ChatGPT~\cite{ChatGPT} in November 2022, large numbers of people have begun to use LLM-based systems as CSTs, consulting LLMs for ideas in creative contexts as wide-ranging as recipe creation~\cite{RecipeCreation}, tabletop roleplaying game scenario design~\cite{TTRPGScenarioDesign}, and marketing slogan generation~\cite{SloganGeneration}.

The adoption of LLMs in creative contexts has raised questions about the extent to which these models can assist in the production of genuinely creative outputs. In particular, some researchers have expressed concern that the widespread use of a small number of highly centralized, data-driven AI systems (such as ChatGPT) may lead to decreased diversity in the outputs of creative processes that incorporate these tools~\cite{epstein2023art,ModelIsTheMessage,AIMediated,PotentialPitfalls}. These concerns resonate with earlier work that has proposed diversity of output as a potential evaluation criterion for AI-based CSTs in general~\cite{ERaCA} and LLM-based CSTs in particular~\cite{HowNovelistsUseLLMs}. Despite these earlier discussions, however, it is only very recently that researchers have begun to directly study the question of whether the use of AI-based CSTs leads to homogenization of human creative output~\cite{PredictableWriting,ERaCA,OpinionatedLanguageModels,PadmakumarHe,DoshiHauser}.

To advance understanding of homogenization effects in human-AI co-creative contexts, we conducted a 36-participant comparative user study of ChatGPT and an alternative, non-AI CST. Participants completed four divergent ideation tasks based on a subset of the Torrance Tests of Creative Thinking (TTCT)~\cite{TTCT}; 
each participant completed half of these tasks with ChatGPT and half with the non-AI CST. Participants produced 1271 ideas in total. Based on the resulting data, we investigate four research questions:

\begin{enumerate}
\item[RQ1] With which CST do participants produce more semantically similar ideas at the \emph{group} level? (A: \textbf{ChatGPT})
\item[RQ2] With which CST do participants produce more semantically similar ideas at the \emph{individual} level? (A: \textbf{no difference})
\item[RQ3] Do ChatGPT users feel more or less \emph{responsible} for the ideas they produce? (A: \textbf{less responsible})
\item[RQ4] Beyond originality, do ChatGPT and non-ChatGPT users differ in terms of \emph{other facets} of creativity, such as fluency, flexibility, and elaboration? (A: \textbf{ChatGPT $\rightarrow$ higher fluency, flexibility, and elaboration})
\end{enumerate}

In the remainder of this paper, we first summarize prior work on the creative homogenization effects of AI-based CSTs, reasons we might expect LLM-based CSTs to result in creative homogenizaton, and approaches to the evaluation of CSTs in general. We then describe our experimental procedure and results. Finally, we discuss potential implications of our findings for users, designers, and developers of LLM-based CSTs. 
Uniquely among recent studies, we compare the homogenization effects of LLMs to those of a potential alternative CST; tease apart individual user-level from group-level homogenization effects; and extend the study of LLM-driven homogenization effects outside the domain of writing.

Collectively, our results suggest that LLM-driven homogenization stems from the LLM providing different users with similar ideas, rather than by increasing individual user-level fixation; that low inferential distance between LLM outputs and apparently finished creative products may contribute to homogenization; and that users may be able to resist homogenization effects if they are given a sense of what the model tends to suggest in similar contexts. These conclusions imply potential mitigations for homogenization effects at the CST design level and clear research directions for follow-up work.

\section{Background and Related Work}
\subsection{Prior Studies of Homogenization}
To date, there have been relatively few direct studies of creative homogenization resulting from the use of AI-based CSTs. Prior to the widespread adoption of LLMs, \citeauthor{PredictableWriting} found that ``predictive text encourages predictable writing''~\cite{PredictableWriting} in the context of single-word suggestions given by smartphone keyboards. Subsequently, \citeauthor{ERaCA} evaluated whether an AI-based poetry composition tool caused users to produce more or less similar poems over time as they continued to use the tool~\cite{ERaCA}.

Two recent user studies of LLMs provide direct evidence of a homogenization effect in LLM-supported writing. \citeauthor{PadmakumarHe} evaluate the baseline GPT-3 model~\cite{GPT-3} versus the instruction-finetuned variant InstructGPT~\cite{InstructGPT} in the context of a short-form argumentative essay writing task and find a homogenization effect from LLM assistance at both the lexical and content levels, but only for the instruction-tuned LLM~\cite{PadmakumarHe}. Meanwhile, \citeauthor{DoshiHauser} evaluate GPT-4 in the context of short-form fictional narrative writing and observe a similar homogenization effect~\cite{DoshiHauser}.

While not addressing homogenization effects directly, several other recent studies examine how LLMs change the writing of humans who use them for writing support~\cite{DSIIWA}. \citeauthor{CoAuthor}~\cite{CoAuthor} find that LLM support tends to increase the diversity of a user's vocabulary but may reduce their feelings of ownership for the text they produce. \citeauthor{OpinionatedLanguageModels}~\cite{OpinionatedLanguageModels} find that an opinionated LLM-based CST influences the opinions expressed by its users in argumentative writing. Similarly, \citeauthor{NextPhraseSuggestions}~\cite{NextPhraseSuggestions} find that LLM-supplied next-phrase suggestions may alter the form and content of a human user's writing even when the user dislikes these suggestions. Potentially explaining these results, \citeauthor{InspirationThruObservation}~\cite{InspirationThruObservation} proposes and provides evidence for an ``inspiration through observation'' model of how human writers are influenced by LLMs: observation of LLM-generated text shapes the ideas expressed by the LLM's user even when the user does not incorporate LLM-generated text directly into their writing. 

Still other studies investigate the homogeneity of LLM outputs in creative contexts without examining their effects on human-in-the-loop creative processes (where a human user is more actively involved in curating and refining LLM output). 
\citeauthor{Begus}~\cite{Begus} compares human-written to GPT-generated short stories, observing that GPT-generated stories are less diverse than human-written stories in structure 
but exhibit greater gender and sexual diversity in character description. In a similar vein, \citeauthor{TTCW}~\cite{TTCW} compare LLM-generated short stories to stories by expert human writers and find that the human-written stories substantially outperform the LLM-generated stories on all dimensions of creativity measured by the TTCT, including the originality dimension.

Altogether, we are aware of only two previous direct studies of LLM-driven creative homogenization. Both of these studies only compare LLM-assisted to tool-unassisted users, making it difficult to determine how the homogenization effects of LLMs compare to those of potential alternative CSTs. Additionally, both studies focus specifically on writing tasks, and consequently both studies solicited only one creative product (i.e., written essay or short story) from each participant; thus, it is unclear whether homogenization effects come into play predominantly at the individual user or group level. Existing evidence seems to support the existence of homogenization effects overall, but more research is needed to gauge the severity of these effects; to confirm their existence outside of writing tasks specifically; and to clarify how they emerge.

\subsection{Reasons to Expect Homogenization}
Concerns about LLM-driven homogenization of creative outputs~\cite{ModelIsTheMessage,epstein2023art} have so far mostly been articulated in terms of \emph{algorithmic monoculture}~\cite{AlgorithmicMonoculture}. The use of a single consistent AI system to perform tasks that used to be performed by many very different humans or systems can lead to increased homogenization of outcomes~\cite{OutcomeHomogenization}; similar effects might be expected on creative processes if many different people all begin using ChatGPT (or another singular LLM-based system) as a CST.

Beyond monoculture, we might also expect LLM-based CSTs to result in homogenization for several other reasons. Creative design processes are subject to measurable \emph{fixation} effects~\cite{DesignFixation,FixationReviewGero,FixationReview2019}, through which the final proposed solutions to a design problem are inappropriately constrained by features of earlier solution candidates~\cite{FixationReview2018}; LLMs may induce fixation by presenting users with complete-seeming ideas early in the ideation process, thereby reducing variation in later ideas. Past work in HCI has proposed that underdetermination may be valuable in human-machine co-creation~\cite{Underdetermination}; by producing text that looks ``finished'', LLM-based CSTs may foreclose the space of desirable ambiguity~\cite{Ambiguity} that results in legitimate creativity. If people trust LLMs due to \emph{trustworthiness cues}~\cite{TrustworthinessCues} such as an authoritative-sounding writing style, or due to a general trust in machines~\cite{MachineHeuristic}, they may treat LLM-suggested ideas as good or valuable by default, causing them to take up these ideas without closely examining them. Further, groups tend to be less creative than individuals when majoritarian convergence processes are used, and minority dissent within group creative processes tends to lead to greater creativity~\cite{DivergenceConvergenceDissent}; an LLM trained to reproduce statistically likely results may seem to speak with the authority of the majority and therefore reduce creativity. 
If LLMs indeed exert homogenization effects on creative processes, any of these proposed mechanisms might be at work.

\subsection{Evaluating Creativity and CSTs}
Evaluation of CSTs has remained an open problem since essentially the beginning of CST research~\cite{EvaluatingCSTsWorkshopReport}, due in part to the ambiguous and multifaceted nature of ``creativity'' as a phenomenon, in part to the wide range of (sometimes contradictory) user needs associated with different creative contexts, and in part to the lack of a clear consensus around what aspects of CSTs should be evaluated~\cite{EvaluatingCSTs}. Broadly speaking, approaches to the evaluation of CSTs can be divided into two categories: those that primarily evaluate aspects of the creative \emph{process} when the CST is used, and those that primarily evaluate the creative \emph{products} that emerge from this process.

On the process side, CSTs are most frequently evaluated by means of subjective self-reports of experience from tool users. The Creativity Support Index (CSI)~\cite{CSI} is a widely used and psychometrically validated survey instrument that attempts to standardize some aspects of this experience reporting process across different CSTs. Other (often bespoke) survey instruments are also deployed in CST evaluation, either as a supplement or an alternative to the CSI (e.g., \cite{MoraiMaker,Germinate,LooseEnds,FlatMagic,PromptPaint}). Process is sometimes also evaluated via observation of user \emph{actions} during the creative process (e.g., \cite{DrawingApprenticeInteractionDynamics,Germinate,ERaCA,DesignStyleClustering,FlatMagic}). Evaluations of LLM-based CSTs have largely followed this pattern to date: most such CSTs are evaluated primarily through subjective experience reports and secondarily through observation of user actions (e.g., \cite{HowNovelistsUseLLMs,StolenElephant,SecondMind,Luminate,ORIBA}).

On the product side, CSTs can also be evaluated by examining the quantity, quality, or other characteristics of the artifacts that their users produce. The Torrance Tests of Creative Thinking (TTCT)~\cite{TTCT} evaluate the creativity of test-takers according to four facets of creative output: \emph{fluency}, or sheer quantity of artifacts created; \emph{flexibility}, or quantity of distinct categories of artifacts created; \emph{originality}, or dissimilarity of created artifacts to others' creations; and \emph{elaboration}, or level of detail in created artifacts. These same criteria can also be applied to the evaluation of CSTs by comparing a novel tool's users to users of another baseline tool along these lines. Notably, the TTCT does not include artifact \emph{quality} as an evaluation criterion; where studies of CSTs have attempted to evaluate output quality, human raters have usually been employed to judge the results (e.g., \cite{Filmmaking,Metaphoria}). This pattern holds for studies of LLM-based CSTs as well (e.g., \cite{TaleBrush,Dramatron,DoshiHauser}).


Our comparison of ChatGPT to a non-AI CST makes use of both process and product data, with a particular focus on the assessment of homogenization effects via examination of creative products: the ideas that study participants produce. Homogenization effects are most closely linked to the \emph{originality} dimension of the TTCT; like other parallel studies of homogenization effects~\cite{PadmakumarHe,DoshiHauser}, we investigate originality primarily by means of semantic similarity, using a well-performing sentence embedding model~\cite{SentenceTransformers} whose direct predecessors~\cite{GloVe} have been found to agree well with human judgments of originality in creativity research~\cite{SemDis,Dumas}. We also use product data to assess CST effects on fluency, flexibility, and elaboration (the other three facets of creativity that the TTCT attempts to gauge) via simple idea count, manual categorization of ideas, and stoplisted word count~\cite{MeasuringElaboration} respectively. Self-reported user experience data (collected via both the CSI and bespoke survey items) is used to assess CST effects on participant feelings of responsibility and other user experience qualities.

\section{Methods}
We conducted a within-subjects experiment to evaluate the effects of using two different CSTs for idea generation: ChatGPT and the Oblique Strategies (OS) deck.

\subsection{Participants}
\subsubsection{Recruitment}
Participants were recruited from academic mailing lists, forums, and solicitations posted by the experimenters on social media. Our study protocol and recruitment materials were approved by the Santa Clara University IRB. Participation was incentivized with a \$17.50 gift card for a one hour session. Participants were required to have a stable Internet connection, a device capable of screen-sharing, and access to a quiet place for the duration of the session in order to participate in the study. Of the 36 participants originally recruited, three were excluded from all analyses for failure to follow direction (e.g., not using ChatGPT when prompted to do so).

\subsubsection{Demographics}
Our sample included 33 participants, ranging in age from 22 to 44 (M=28.36, SD=6.84), and including 63.63\% (n=21) men and 33.36\% (n=12) women. We had 39.39\% Black or African American (n=13), 36.36\% Asian (n=12), and 24.24\% (n=8) White participants. Participant occupations included 36.36\% students (n=12), 30.30\% creative professionals (e.g. game designer, writers, n=10), and 33.33\% other professionals (e.g registered nurse, social worker, customer service, n=11). Educational experience included 15.15\% high school graduates (n=5), 15.15\% participants with some college (n=5), 48.48\% college graduates (n=16), and 21.21\% participants with a Master's degree or higher (n=7). Experience with text-based generative AI (e.g. ChatGPT) was varied, with 72.72\% of participants reporting daily (n=16) or weekly (n=8) usage of LLM tools, and 27.27\% of participants reporting that they had used them once a month or less (n=3), or that they had never used them before the study (n=6). Experience with generative art AI tools (e.g. Midjourney, Stable Diffusion) was less common: 63.63\% of participants reported that they had either never used such tools (n=13) or used them once a month or less (n=9), and 33.33\% of participants reported that they used them about once a week (n=5) or daily (n=6). 

\subsection{Materials}
\subsubsection{ChatGPT}
ChatGPT is a popular LLM-based tool trained to respond to text instructions \cite{ChatGPT}. Participants in this study used the versions of ChatGPT 3.5 released on May 3rd 2023 (n=6, 16.6\%) and on August 3rd, 2023 (30, 83.3\%).

\subsubsection{Oblique Strategies Deck}
The Oblique Strategies (OS) deck, originally created by the artists Brian Eno and Peter Schmidt \cite{eno1975oblique}, consists of a collection of cards with prompts designed to support creative work. Example prompts include ``Turn it upside down'', ``Don't avoid what is easy'', ``Destroy the most important thing'', and ``How would someone else do it?''  We directed participants to a web app version of the deck \cite{ruten_oblique_2012} as an alternative CST to ChatGPT in our control condition.

\subsubsection{Creative Ideation Prompts}
We provided creative ideation prompts for two types of divergent thinking tasks: \emph{Product Improvement} (PI) and \emph{Improbable Consequences} (IC). Prompts included:
\begin{itemize}
\item How could you make a stuffed toy animal more fun to play with? (PI\_A)
\item How could you make a jigsaw picture puzzle more interesting and engaging? (PI\_B)
\item Suppose that a great fog has fallen over the earth and all we can see of people is their feet. What would happen? (IC\_A)
\item Suppose that gravity suddenly became incredibly weak, and objects could float away easily. What would happen? (IC\_B)
\end{itemize}
For each prompt, participants were instructed to generate as many ideas as possible and to try to come up with ideas that no one else would think of.

\subsubsection{Creativity Support Index}
The Creativity Support Index (CSI) is a survey instrument for assessing the ability of a CST to assist a user engaged in creative work \cite{CSI}. Its design was inspired by the NASA TLX, a questionnaire designed to evaluate mental task load \cite{NASATLX}. We administered the CSI to capture participant experiences with each CST after they used that CST to complete a creative ideation task.

\subsection{Procedure}
All experimental sessions were remote-moderated over videoconferencing software. In each session, participants were asked to generate ideas in response to several ideation prompts, first while using one of two CSTs (ChatGPT or OS) and then while using the other tool. Participants were instructed to generate as many ideas as they could, and to try to come up with ideas that no one else would think of. 

During the session, participants were encouraged to think-aloud, to the degree that they felt doing so would not interfere with their performance. The time was held constant at 8 minutes per ideation prompt, and participants responded to two prompts with each support tool. With each tool, the first prompt asked participants to come up with ideas for improving an existing product (Product Improvement). The second prompt asked them to consider an impossible situation and imagine as many possible consequences as they could think of (Improbable Consequences). The order of CSTs and prompts was randomized per participant and balanced across the entire experiment. 

After using each tool the participants responded to the Creativity Support Index questionnaire, and indicated the degree to which they felt personally responsible for their output, or that they felt their output came from the tool that they used. Each session concluded with an open-ended discussion of each participant's experience with both ChatGPT and OS.

\section{Results}
We observed three categories of data about working with each CST: creative process, creative outcomes, and retrospective reflections on the experience of working with the tool. Creative process data included video-recordings of moderated sessions conducted with a talk-aloud protocol, which resulted in observations about prompting styles and iteration, and how participants used ChatGPT output. Creative outcomes included the list of ideas generated with each CST. Retrospective reflections included Creativity Support Index (CSI) ratings, ratings of how personally responsible participants felt for the ideas they generated (vs. crediting the CST), and a brief open-ended discussions about the experience of using each CST conducted at the end of each experimental session. The results reported below are organized by research question, addressing homogenization (at group and individual levels), participants' sense of responsibility for the ideas they produced, and other facets of creativity. We also provide a summary of observations from participant interviews, and some observations of their creative process. Data from three participants who did not follow directions (e.g., not using ChatGPT when instructed to do so) were excluded from our analysis. 

\subsection{Homogenization}
\subsubsection{Evaluating Homogenization via Semantic Similarity}
To evaluate creative outcomes we followed two approaches to quantifying participant-generated ideas, complementing a traditional human rater-based approach with semantic similarity assessment via sentence embeddings~\cite{SentenceTransformers}. Sentence embeddings allow us to quantify homogenization in the form of semantic similarity, comparing the cosine similarity of each idea a participant generated to the average embedding of all participant ideas, to evaluate how distinct an individual participant's ideas are from the group. By comparing to an average embedding for the individual, rather than the group, we are also able to evaluate the diversity of each participant's ideas. 

Our semantic similarity-based approach to homogenization analysis closely follows recent psychological studies of creativity---e.g., \cite{SemDis}. However, the specific sentence embeddings that we used for our homogenization analysis (though high-performing in general) have not previously been validated for creativity assessment. As a result, we performed a small experiment to validate the agreement of these embeddings with human judgments of semantic similarity on our dataset. This experiment is discussed in Appendix \ref{sec:ValidatingEmbeddings}.

\subsubsection{Group-Level Homogenization}
\begin{figure}
    \centering
    \includegraphics[width=\columnwidth]{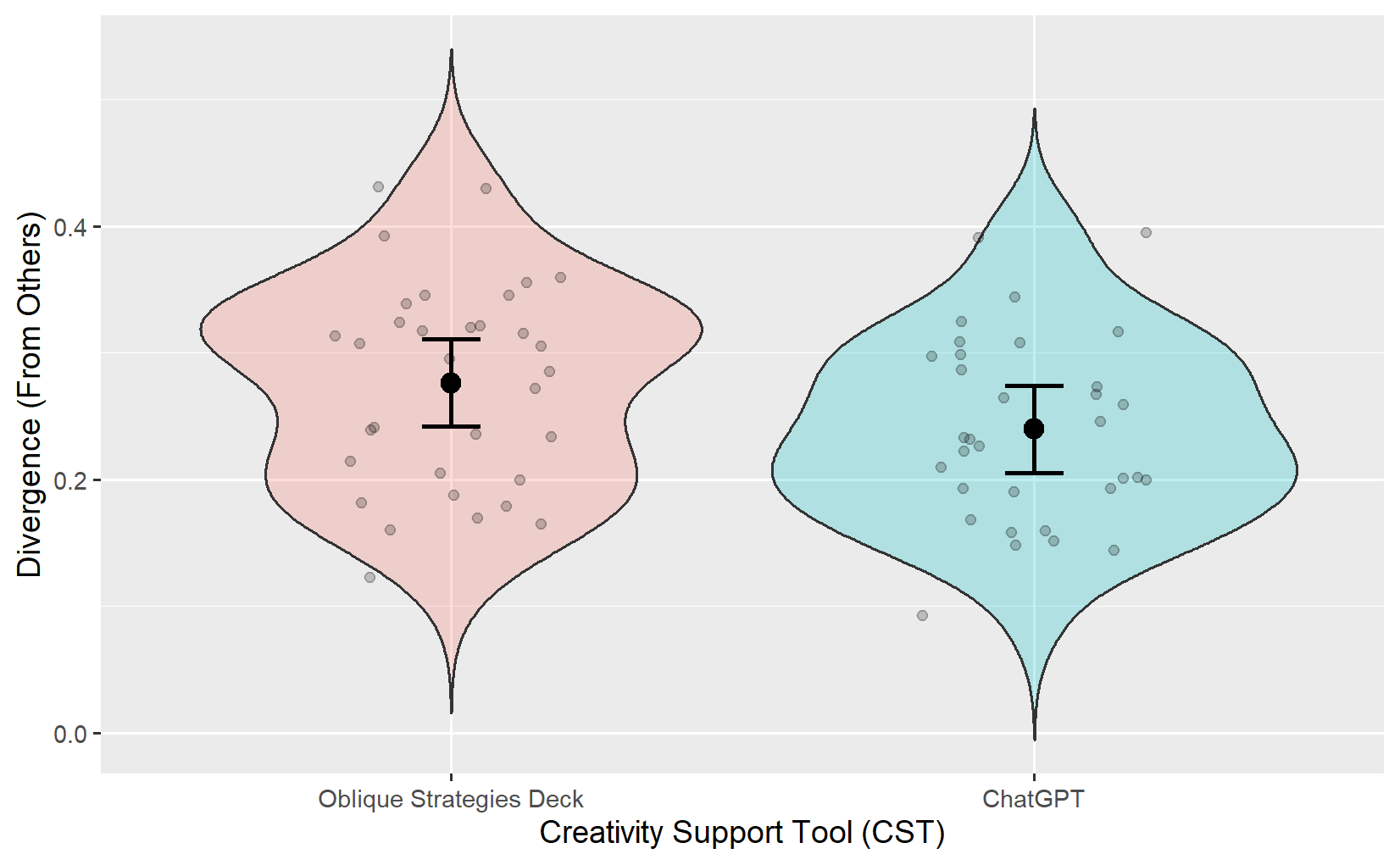}
    \caption[LOD]{Participant responses were more homogenous at the group level (i.e., more semantically similar to the average embedding of \emph{all} participant ideas) when using ChatGPT.\protect\footnotemark[1]}
    \label{fig:Divergence}
\end{figure}

When participants used ChatGPT, the ideas they produced were less divergent from the average embedding of all ideas generated for that task, (\textit{M} = .24, \textit{SD} = .07), compared to the ideas that they produced when using OS, (\textit{M} = .28, \textit{SD} = .08), \textit{t}(32) = 2.154, \textit{p} = 0.038, \textit{d} = .47, 95\% CI [.00,.07]. See Figure \ref{fig:Divergence}. At the group level, ideas produced with the help of ChatGPT were more homogenized.
\footnotetext[1]{All error bars are 95\% Confidence Intervals, with Cousineau-Morey (2008) corrections for within-subjects data \cite{morey2008confidence}.}

\subsubsection{Individual-Level Homogenization}
\begin{figure}
    \centering
    \includegraphics[width=\columnwidth]{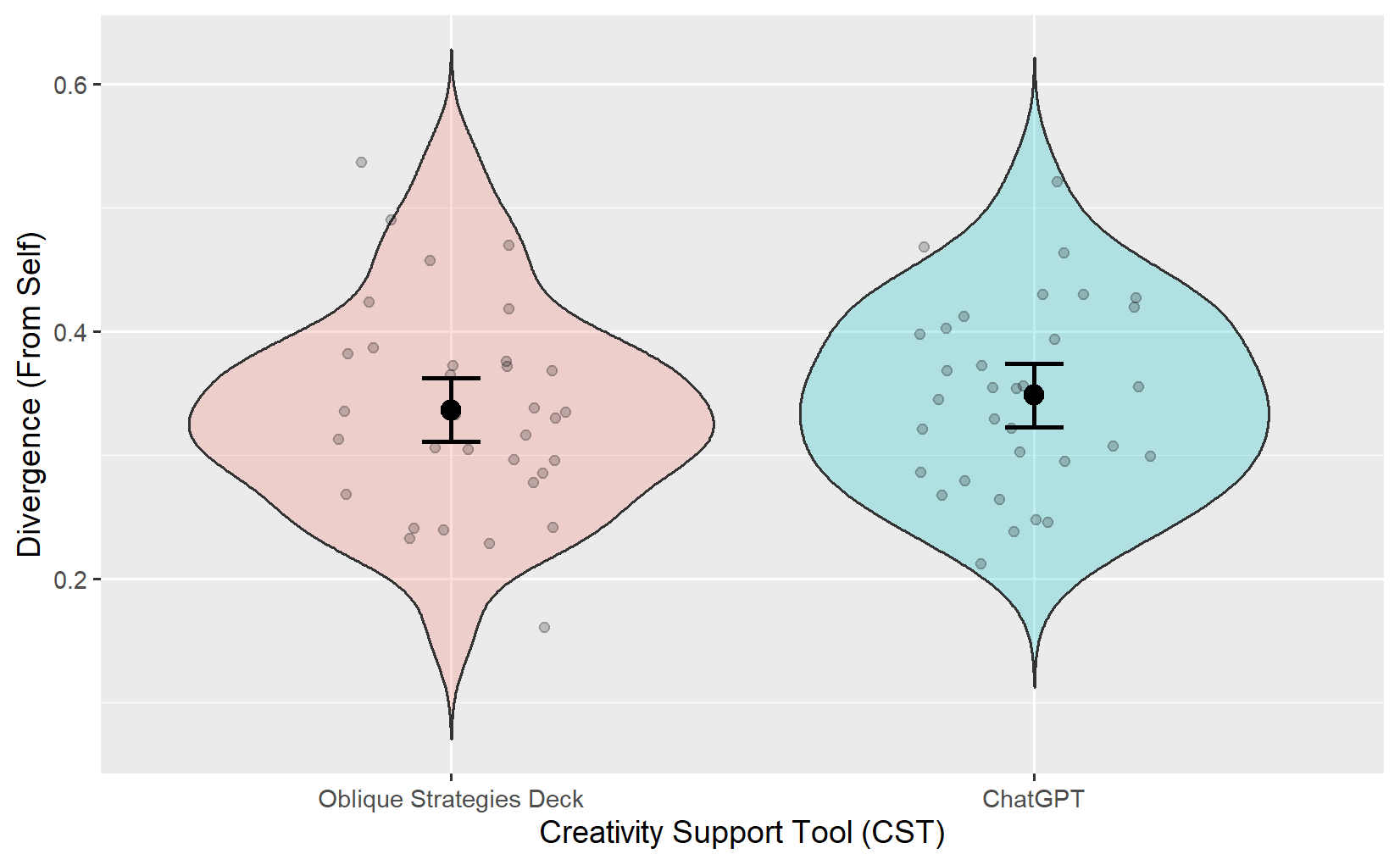}
    \caption{Participant responses were not observably more or less homogenous at the individual level (i.e., more semantically similar to the average embedding of \emph{this participant's} ideas) when using ChatGPT.}
    \label{fig:DivergenceSelf}
\end{figure}

When participants used ChatGPT, the ideas they produced were not observably more divergent from the average embedding of all of the other ideas that they themselves generated for the same ideation prompt, (\textit{M} = .65, \textit{SD} = .07), compared to semantic dissimilarity for ideas that they produced when using OS, (\textit{M} = .66, \textit{SD} = .08), \textit{t}(32) = .944, \textit{p} = 0.352, \textit{d} = .12, 95\% CI [-.04,.01]. See Figure \ref{fig:DivergenceSelf}. At the individual level, we did not observe a difference in homogenization for ideas generated with the help of ChatGPT.

\subsection{Sense of Responsibility}
\begin{figure}
    \centering
    \includegraphics[width=\columnwidth]{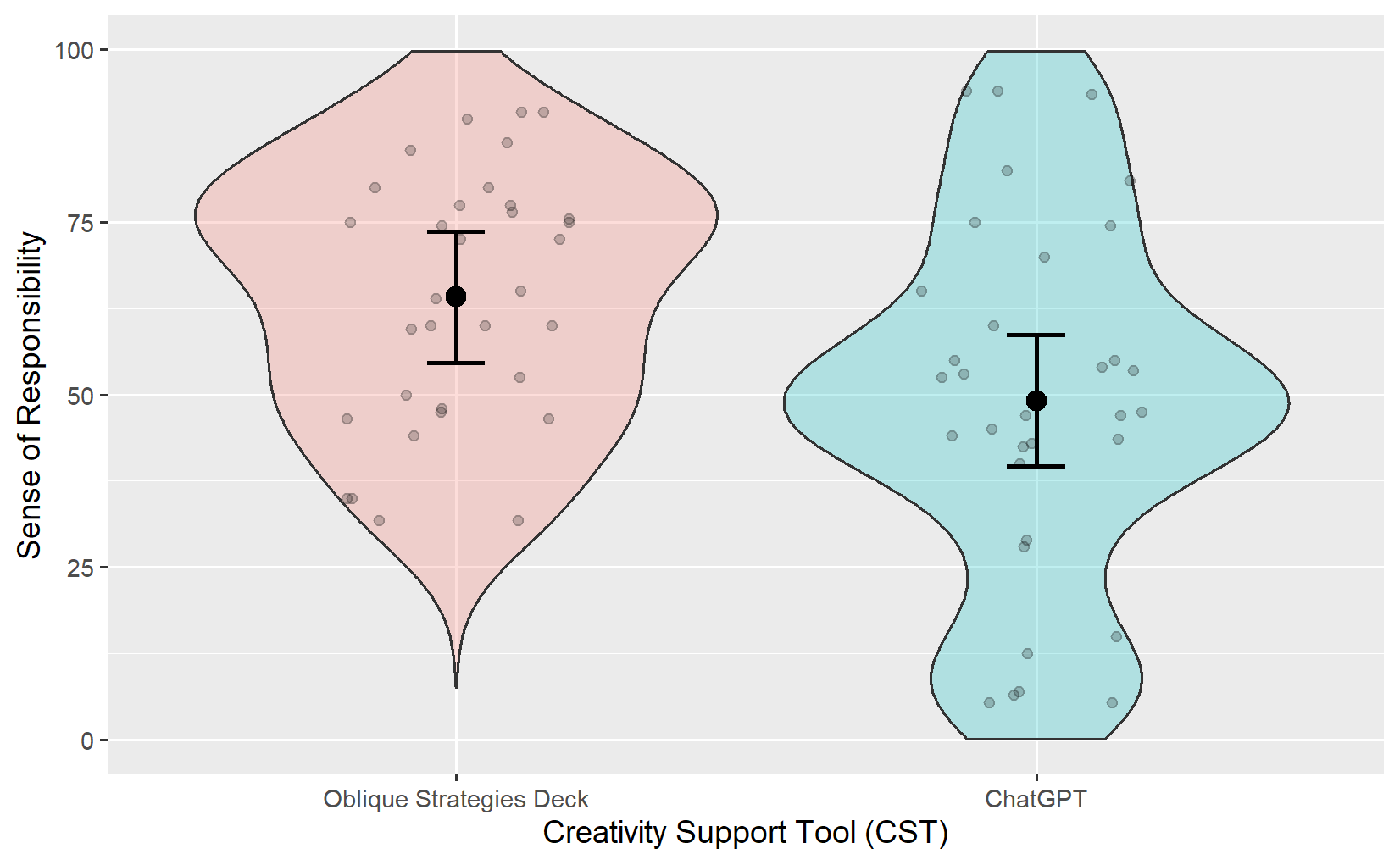}
    \caption{Participants assigned less creative responsibility to themselves, and more to the CST, when working with ChatGPT.}
    \label{fig:Responsibility}
\end{figure}

Participants assigned less responsibility to themselves (and more to the tool) for ideas generated while using ChatGPT (\textit{M} =48.17\%, \textit{SD} =26.22\%), compared to ideas generated while using OS (\textit{M}=63.63\%, \textit{SD} =17.36\%), \textit{t}(32) = 3.21, \textit{p} = 0.003, \textit{d} = .67, 95\% CI [-24.60\%, -5.51\%,]. See Figure \ref{fig:Responsibility}.

\subsection{Other Facets of Creativity}

In addition to the TTCT dimension of originality, quantified in our homogenization analysis above, we also studied the three other TTCT dimensions of creativity quantitatively: fluency by simple idea count, flexibility by human coding of ideas into categories, and elaboration via stoplisted word count~\cite{MeasuringElaboration}. Further, we assessed originality from another angle via several different means of gauging idea uniqueness. We did not evaluate idea quality, which is subjective; may vary substantially depending on the assumed use case for the idea; and is both time-consuming and costly to evaluate via the most widely used method (annotation by human raters).

\subsubsection{Fluency: Simple Idea Count}
\begin{figure}
    \centering
    \includegraphics[width=\columnwidth]{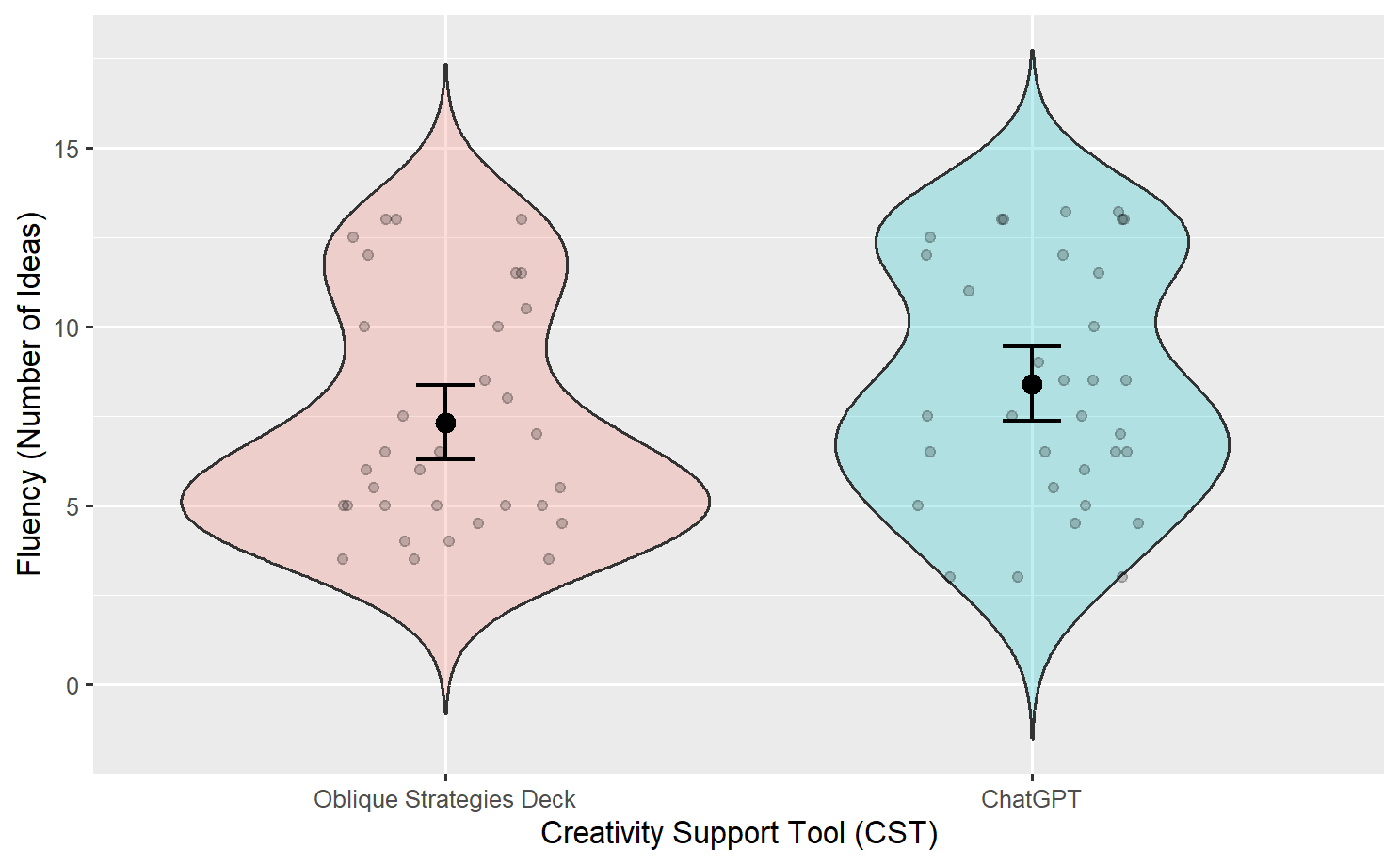}
    \caption{Participants generated more ideas with ChatGPT than with OS.}
    \label{fig:Fluency}
\end{figure}
Participants generated about one additional idea---approximately a 15\% increase---when using ChatGPT (\textit{M} = 8.39, \textit{SD} =3.39), compared to the number of ideas they generated when using OS (\textit{M} = 7.32, \textit{SD} = 3.22), \textit{t}(32) = 2.10, \textit{p} = 0.044, \textit{d} = .32, 95\% CI [.03,2.11]. See Figure \ref{fig:Fluency}.

\subsubsection{Flexibility: Idea Categories}\label{sec:Flexibility}
\begin{figure}
    \centering
    \includegraphics[width=\columnwidth]{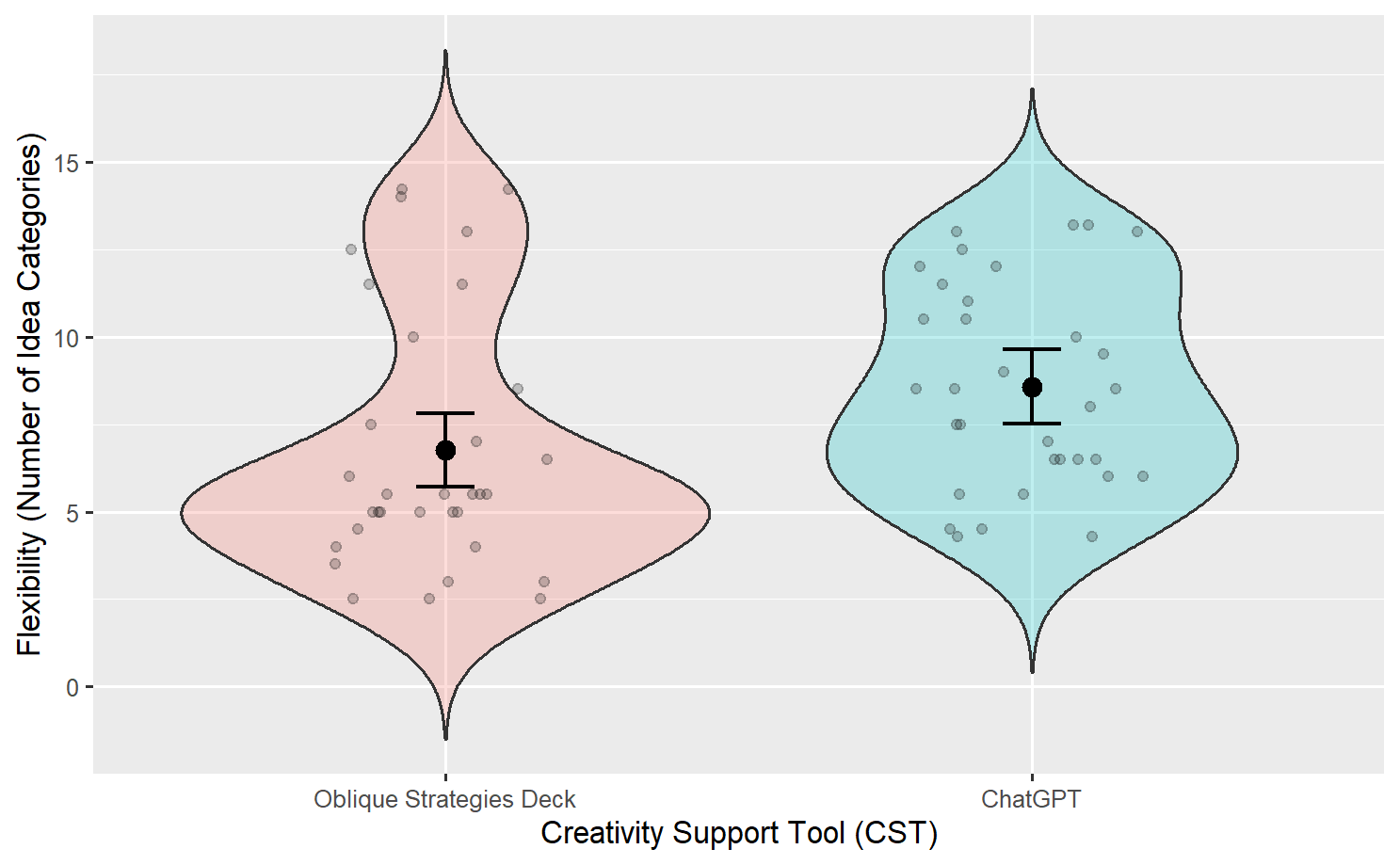}
    \caption{Participants generated ideas across more categories with ChatGPT than with OS.}
    \label{fig:Flexibility}
\end{figure}
Following an iterative grounded theory based approach \cite{charmaz_constructing_2006}, we reviewed the ideas generated for each ideation prompt and observed the categories of ideas that emerged, ultimately generating 181 distinct idea categories from 1271 individual responses. Each participant's responses were then tagged with all relevant idea categories. This process resulted in a count of idea categories hit by each participant, including a sub-count of each participant's idea categories hit that were unique within our sample. Coding was conducted while blind to the CST condition (i.e., coders did not know if an idea was generated with or without LLM support).

Participants generated ideas that hit about 27\% more categories when using ChatGPT (\textit{M} = 8.58, \textit{SD} =2.90), compared to the number of categories covered when using OS (\textit{M} = 6.77, \textit{SD} = 3.69), \textit{t}(32) = 3.50, \textit{p} = 0.001, \textit{d} = .54, 95\% CI [.75,2.86]. See Figure \ref{fig:Flexibility}.

\subsubsection{Elaboration: Stoplisted Word Count}
\begin{figure}
    \centering
    \includegraphics[width=\columnwidth]{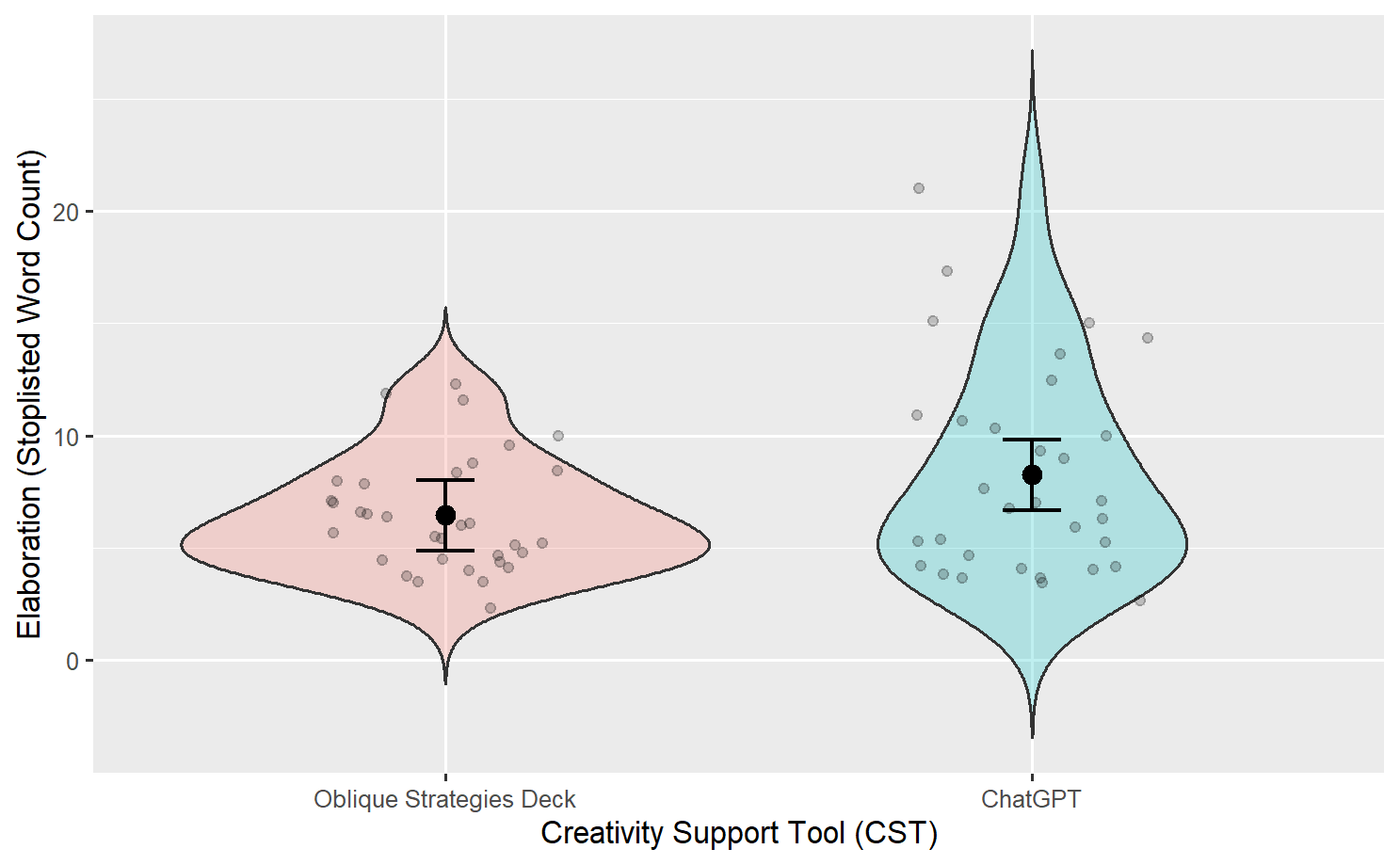}
    \caption{Participants generated more elaborated ideas with ChatGPT than with OS.}
    \label{fig:Elaboration}
\end{figure}
Among several potential computational means of gauging idea elaboration (which is roughly synonymous with complexity or level of detail), stoplisted word count (i.e., counting the words used to express the idea while excluding a ``stoplist'' of common low-information filler words) demonstrates the strongest correlation with human judgments of elaboration~\cite{MeasuringElaboration}. Participant ideas generated while using ChatGPT had a greater stoplisted word count (\textit{M} = 8.25, \textit{SD} = 4.61) than ideas generated while using OS, (\textit{M} = 6.46, \textit{SD} = 2.55), \textit{t}(32) = 2.32, \textit{p} = 0.237, \textit{d} = .48, 95\% CI [.22,3.37], See Figure \ref{fig:Elaboration}.

\subsubsection{Originality: Unique Ideas}
We did not observe a difference in the number of unique ideas (i.e., ideas that no other participant also generated)\footnote[2]{Creativity research sometimes also deals with \emph{historical} uniqueness, analogous to Boden's ``H-creativity''~\cite{Boden}. However, we neither expected nor saw any historically unique ideas submitted by participants in our study.} between participants using ChatGPT (\textit{M} = .74, \textit{SD} = .69) and OS, (\textit{M} = .97, \textit{SD} = 1.00), \textit{t}(32) = 1.65, \textit{p} = 0.108, \textit{d} = .26, 95\% CI [-.05,.50]. 

Because uniqueness is sensitive to sample size \cite{silvia2008assessing}, we also attempted two common alternative approaches to assessing originality: counting idea categories that less than 5\% of the participants produced as unique~\cite{milgram1976creative}, and weighting each idea category by the frequency of its occurrence in the entire sample for this study to produce a weighted flexibility score \cite{runco1987psychometric, guenther2017creativity}.

However, neither of these approaches changed our findings. Applying a 5\% uniqueness threshold to idea categories, we again did not observe a difference in the number of unique responses (with a 5\% threshold) produced by participants using ChatGPT (\textit{M} = .87, \textit{SD} = .88) or OS (\textit{M} = .90, \textit{SD} = 1.07), \textit{t}(32) = .134, \textit{p} = 0.894, \textit{d} = .02, 95\% CI [-.34,.39]. Furthermore, as  measured by our weighted flexibility score, participants using ChatGPT did not generate more unique ideas (\textit{M} = 6.34, \textit{SD} = 3.31) compared to participants using OS, (\textit{M} = 5.50, \textit{SD} = 3.01), \textit{t}(32) = 1.424, \textit{p} = 0.164, \textit{d} = .26, 95\% CI [-.35,2.02].

\subsection{Retrospective Reflections}

\begin{table*}[ht!]
\centering
\resizebox{\textwidth}{!}{%
\begin{tabular}{llrr}
\textbf{Theme} & \textit{\textbf{Example Responses}} & \multicolumn{1}{l}{\textit{\textbf{n}}} & \multicolumn{1}{l}{\textbf{\%}} \\ \hline
\rowcolor[HTML]{EFEFEF} 
\begin{tabular}[c]{@{}l@{}}Effort/Reward Tradeoff (Oblique Strategies)\\ OS was more challenging to use, but also more rewarding.\end{tabular} & \begin{tabular}[c]{@{}l@{}}\textit{Oblique Strategies got me thinking more creatively, but I got more responses with ChatGPT.}\\ \textit{It was harder to use Oblique Strategies, but it was more fun and it got me to more interesting places.}\end{tabular} & \textit{10} & 27.78\% \\
\rowcolor[HTML]{FFFFFF} 
\begin{tabular}[c]{@{}l@{}}Speed/Accuracy (ChatGPT)\\ ChatGPT was fast, and its responses were accurate.\end{tabular} & \begin{tabular}[c]{@{}l@{}}\textit{ChatGPT gave me the right answers.}\\ \textit{Using ChatGPT is a very nice experience... It's very fast and accurate.}\end{tabular} & \textit{9} & 25.00\% \\
\rowcolor[HTML]{EFEFEF} 
\begin{tabular}[c]{@{}l@{}}Low Engagement (ChatGPT)\\ Using ChatGPT was less engaging.\end{tabular} & \begin{tabular}[c]{@{}l@{}}\textit{ChatGPT allowed me to turn my brain off. It did more of the heavy lifting.}\\ \textit{ChatGPT reduced the confidence I had to come up with creative things on my own.}\end{tabular} & \textit{8} & 30.56\% \\
\rowcolor[HTML]{FFFFFF} 
\begin{tabular}[c]{@{}l@{}}Low Task Relevance (Oblique Strategies)\\ Responses from OS were less task-relevant.\end{tabular} & \begin{tabular}[c]{@{}l@{}}\textit{I didn't really understand Oblique Strategies. It didn't relate to most of the questions.}\\ \textit{The cards were inspirational, but most of them were just random thoughts.}\end{tabular} & \textit{7} & 19.44\% \\
\rowcolor[HTML]{EFEFEF} 
\begin{tabular}[c]{@{}l@{}}Repetitive Responses (ChatGPT)\\ ChatGPT responses were repetitive.\end{tabular} & \begin{tabular}[c]{@{}l@{}}\textit{ChatGPT is a more research-based tool. ChatGPT is a bit repetitive, but it has a lot of data.}\\ \textit{When I asked for more {[}ChatGPT{]} repeated half... When I want more, I want different more.}\end{tabular} & \textit{3} & 8.33\% \\
\rowcolor[HTML]{FFFFFF} 
\begin{tabular}[c]{@{}l@{}}High Engagement (Oblique Strategies)\\ OS was more engaging.\end{tabular} & \begin{tabular}[c]{@{}l@{}}\textit{I got into a flow with Oblique Strategies.}\\ {[}\textit{Oblique Strategies cards}{]} \textit{were more interesting than ChatGPT.}\end{tabular} & \textit{3} & 8.33\% \\
\rowcolor[HTML]{EFEFEF} 
\begin{tabular}[c]{@{}l@{}}Premature Closure (ChatGPT)\\ The ChatGPT responses became too specific too quickly.\end{tabular} & \begin{tabular}[c]{@{}l@{}}\textit{ChatGPT feels like it can go really specific really quickly. Almost more than you need.}\\ \textit{With ChatGPT, I felt like it was more guided and way more specific.}\end{tabular} & \textit{2} & 5.56\% \\
\rowcolor[HTML]{FFFFFF} 
\begin{tabular}[c]{@{}l@{}}Self-Doubt (ChatGPT)\\ Self-deprecation regarding technical ability.\end{tabular} & \begin{tabular}[c]{@{}l@{}}\textit{I felt like ChatGPT was a tool I could use in a deeper way.}\\ \textit{I didn't really feel like I knew good questions to ask ChatGPT.}\end{tabular} & \textit{2} & 5.56\% \\ \hline
\end{tabular}%
}
\caption{Reflections on experiences with idea generation using both CSTs (ChatGPT and OS).}
\label{tab:reflections}
\end{table*}

\subsubsection{Interview}
Participants were asked to discuss their own experiences using both tools, and several themes emerged from those discussions. The single most common theme was that ChatGPT was easier to use but less rewarding (27.78\%, \textit{n} = 10). The second most common theme was a positive sentiment regarding the speed and accuracy of the LLM-provided responses (25.00\%, \textit{n} = 9). Participants also reported finding ChatGPT to be less engaging (22.22\%, \textit{n} =8), that the LLM responses were too repetitive (8.33\%, \textit{n} = 3) and that the LLM responses became too specific too quickly (5.56\%, \textit{n} = 2). Participants also made self-deprecating remarks regarding their familiarity and technical ability with ChatGPT (5.56\%, \textit{n} = 2), but made no such remarks regarding OS.

\subsubsection{Creativity Support Index}
We did not observe any differences in Creativity Support Index (CSI) ratings for the LLM-based tool (\textit{M}=78.03\%, \textit{SD} =18.82\%) and for OS (\textit{M}=73.98\%, \textit{SD} =15.35 \%), \textit{t}(35) = 1.028, \textit{p} = 0.312, \textit{d} = .24, 95\% CI [-3.94\%, 12.02\%]. We also observed no differences for any of the CSI sub-scales (\textit{Exploration, Engagement, Effort/Reward Tradeoff, Tool Transparency, Expressiveness}). We did not collect responses for the \textit{Collaboration} subscale, which is irrelevant and often omitted in exclusively single-user contexts like that of our study (e.g., \cite{MultimodalPenBased,GANCollage}).

\subsection{Creative Process}
\subsubsection{Initial Ideation}
When using ChatGPT, about two-thirds (63.89\%, \textit{n}=23) of our participants began by providing a number of their own ideas (\textit{M}=5.65, \textit{SD}=3.90) before interacting with the tool at all. This was similar to the interaction pattern we observed with OS, and contrary to our expectation that participants might rely on the CST to overcome a blank page~\cite{BlankPageParalysis,UnmetNeeds}.

\subsubsection{Prompting Styles and Iteration}
When interacting with ChatGPT, the majority of participants (86.11\%, \textit{n} = 31) directly copy/pasted the creative ideation prompt for their current task. Other prompting strategies included asking for relevant scientific or factual information (e.g. ``What are the most popular dolls?'', ``How common is face blindness?'', etc.), asking for responses from a specific perspective (e.g. project manager, scientist, athlete, etc.), or adding creative constraints (e.g. make this toy for a dog, respond in the form of a story, etc.). Most participants (72.22\%, \textit{n} = 26) also iterated on their prompt, either by adding more context to their initial prompt, probing with more specific questions or variations, or by asking ChatGPT to regenerate answers or provide additional responses. 

\subsubsection{ChatGPT Output Usage}
When using ideas from ChatGPT, a slight plurality of participants copy and pasted from the output directly (41.67\%, \textit{n} = 15). The majority of these participants copied selectively: only 1 copied the entirety of the ChatGPT output without any attempt at curation or editing. A minority of participants entirely avoided direct copying, providing ideas either paraphrased from or inspired by the ChatGPT output (16.67\%, \textit{n} = 6). A substantial fraction of participants combined both approaches, copying a few ideas directly, modifying others, and later adding more of their own ideas with no obvious connection to the ChatGPT output (38.89\%, \textit{n} = 14).

\subsubsection{Process Impact on Outcome}
\label{sec:ProcessImpact}
We observed a significant positive relationship between the number of prompts a participant entered into ChatGPT and the number of ideas that they generated, \textit{r}(32)=.43, \textit{p} = .008, and between the number of prompts entered and weighted uniqueness scores, \textit{r}(32)=.46, \textit{p} = .004. We did not observe any relationship between the number of LLM prompts entered and average homogeneity of ideas generated, \textit{r}(32)=.23, \textit{p} = .185.

We did not observe any difference in the number of ideas generated from participants who started by entering their own ideas first (\textit{M} = 8.38, \textit{SD} = 3.39), compared to those who went directly to ChatGPT, (\textit{M} = 8.71, \textit{SD} = 3.58), \textit{t}(32) = .276, \textit{p} = .784, 95\% CI [-2.11,2.77]. We also observed no difference in the weighted number of unique ideas (Own Ideas First: \textit{M} = 7.21, \textit{SD} = 3.12, LLM First: \textit{M} = 7.46, \textit{SD} = 3.12), \textit{t}(32) = .222, \textit{p} = .826,  95\% CI [-1.98,2.47], or in the homogeneity of their ideas, between participants who took these different approaches, (Own Ideas First: \textit{M} = .99, \textit{SD} = .03, LLM First: \textit{M} = .99, \textit{SD} = .03), \textit{t}(32) = .208, \textit{p} = .836, 95\% CI [-.01,.02]. 

We did not observe any difference in the number of ideas generated from participants who used any creative prompting strategy (e.g. role-based, creative constraints, etc.) (\textit{M} = 9.18, \textit{SD} = 3.77), compared to those who directly copied the creative ideation prompt, (\textit{M} = 8.01, \textit{SD} = 3.13), \textit{t}(32) = 1.016, \textit{p} = 0.317, 95\% CI [-3.51,.1.17]. We also observed no difference in the weighted number of unique ideas, (Unusual Prompt: \textit{M} = 7.98, \textit{SD} = 3.12, Copied Prompt: \textit{M} = 6.82, \textit{SD} = 3.12), \textit{p} = .278, 95\% CI [-3.29,.98], or in the homogeneity of their ideas between participants who took these different approaches, (Unusual Prompt: \textit{M} = .99, \textit{SD} = .03, Copied Prompt: \textit{M} = .99, \textit{SD} = .03), \textit{t}(32) = .091, \textit{p} = 0.927, 95\% CI [-.02,.02]. 

\section{Discussion}
Overall, we found that---in line with to our expectation of LLM-induced homogenization---ideas generated with assistance from ChatGPT were significantly less semantically diverse at the group level than ideas generated with assistance from the non-AI-based CST. When supported by ChatGPT, participants produced a larger number of ideas, but the set of ideas generated by each individual participant was similarly diverse in the ChatGPT and non-ChatGPT conditions, suggesting that the increase in quantity of ideas generated did not result in a commensurate increase in diversity as might be expected if diversity is a linear function of quantity~\cite{simonton2010creative}.

We interpret our results as supporting three key takeaways. First, we find that homogenization stems not from individual-level increases in fixation when working with the LLM, but from group-level suggestion of similar ideas to different users by the LLM. Second, we suggest that the homogenization effect of LLMs on creative ideation is attributable in part to the low inferential distance between LLM outputs and apparently complete ideas. Third, we infer that LLM users might be able to resist the homogenization effect if they are given information about what kinds of stereotyped outputs the LLM tends to produce in a particular creative context, though it may still be difficult for them to adapt their prompting strategies to elicit more diverse responses from the LLM. We discuss these and several other takeaways below.

\subsection{Creative Fixation}
Creative fixation~\cite{DesignFixation,FixationReviewGero,FixationReview2018,FixationReview2019} occurs when people engaged in a creative activity become incapable of seeing past an inappropriately narrow assumption that they have made about the conceptual space in which they are working. If LLMs induced creative fixation (for instance by presenting users with ideas to which they cannot readily envision compelling alternatives), we would expect to see individual-level homogenization brought about by fixation: decreased diversity within the set of ideas proposed by each individual user, due to users becoming fixated on specific subsets of the space of all possible ideas. We did not witness this in practice, suggesting that LLMs do not increase homogenization by causing or worsening creative fixation.

This remains the case despite the fact that some participants in our study viewed ChatGPT as a ``very fast and accurate'' authoritative source that ``gave [...] the right answers'', a view which might be expected to provoke user fixation on ideas suggested by the LLM. Even among participants who viewed ChatGPT as an authority, they still generally suggested some ideas unlike those generated by the LLM, perhaps recognizing (as some users explicitly noted) that the LLM tended to ``go really specific really quickly'' or become ``repetitive'' in its outputs. Altogether, we conclude that users viewing ChatGPT as authoritative may have led them to \emph{accept} LLM-suggested ideas as valid, but it did not seem to \emph{constrain} the process of ideation by displacing ideas that were somehow incompatible with those suggested by ChatGPT.

\subsection{Inferential Distance}
In planning this study, we considered it a possibility that the Oblique Strategies (OS) deck might cause greater homogenization than ChatGPT due to the fixed nature of the OS cards. The deck consists of a relatively small set of fixed text strings, and if different users drew the same card in the context of the same ideation prompt, we expected that the similar stimulus might push them to think in similar directions. In practice, however, this did not turn out to be the case. We believe that this might be due to higher \emph{inferential distance}~\cite{InferentialDistance} between the text on the OS cards and the expected form of the task responses: because the user has to do more work to interpret the text of a particular OS card in relation to a particular ideation prompt (compared to text generated by ChatGPT in response to that prompt), the OS deck creates more room within the overall ideation process for the user's psychological uniqueness to influence the eventual output ideas.

Subjective sense of responsibility data further substantiate this interpretation. Compared to Oblique Strategies users, ChatGPT users reported feeling less responsible for the ideas they produced, and several mentioned their own felt sense of non-responsibility for ideas produced with ChatGPT in post-test interview responses. In the words of one participant, ``this was a creativity task, and it felt like I should have some stake in the answer''. For at least some users, the apparent completeness of ChatGPT responses resulted in users feeling disengaged from the process of idea generation; according to another participant, ``ChatGPT allowed me to turn my brain off'' and did ``the heavy lifting'' during ideation. However, we note that even disengaged users typically did not simply copy-paste ChatGPT output unexamined into their list of ideas: we only observed one participant in the entire study directly copy-paste the entirety of a ChatGPT response without any further curation or modification, suggesting that even low-engagement users may be experiencing something closer to ``inspiration through observation''~\cite{InspirationThruObservation} than to total subsumption of the ideation process by the CST.

From one perspective, low inferential distance between CST outputs and finished-looking artifacts can be viewed as enabling \emph{algorithmic loafing}~\cite{AlgorithmicLoafing}: simply accepting and passing along the decisions made by the algorithm, regardless of whether these decisions agree with the decisions the user might have made without the algorithm's input. Consequently, one way to mitigate the homogenization effect of LLM-based CSTs may be to design the CST to output deliberately oblique or gnomic responses, analogous to those printed on Oblique Strategies cards: outputs that can provoke user ideation in potentially unexpected directions, but cannot be employed as straightforward substitutes for user ideas (instead requiring a degree of user interpretation before they can be used). In this way, future LLM-based CSTs may be able to reintroduce some of the underdetermination~\cite{Underdetermination} that has proven valuable in other computationally engaged creative processes.

\subsection{User Adaptation to Homogenization}

The absence of individual-level homogenization in our study suggests that participants were able to gauge the diversity of their own responses and continue producing ideas until they felt the set of ideas they have produced is sufficiently diverse. This may be because participants can see all of their own ideas, so they can actively modulate the diversity of the set of ideas that they have generated---for instance by generating new ideas (either on their own, by prompting ChatGPT and curating the output, or both) until they are satisfied with the level of diversity in the resulting idea set overall. However, participants cannot easily gauge the similarity or dissimilarity of their own responses to other people's responses, so they cannot actively intervene to modulate diversity at the group level. The ability of participants to observe and work around homogenization at the individual level is backed up to some extent by the fact that a few ChatGPT users explicitly noted the repetitiveness of the model's responses. ``When I want more'', one user asserts, ``I want \emph{different} more.''

The apparent success of individual-level user adaptation to homogenization suggests that one LLM-based CST design strategy for mitigating homogenization may involve detecting and alerting the user to clich\'es within model output, as has been suggested before~\cite{ReflectiveCreators}. By making it apparent to users when a particular piece of model output is semantically similar to outputs that other users have previously encountered, it may be possible to give users more information about the typicality of model output, and thereby to help them notice and resist model-induced homogenization at the group level as well. 

More generally, users may also be able to adapt to homogenization in the real world via adoption of and horizontal movement between a variety of different CSTs, each of which imposes a different ``normative ground'' on users and their creative inputs~\cite{CSTsAndPower}. To mitigate homogenization, CST designers might thus seek to create a plurality of different tools between which users can move, rather than a small number of monolithic, one-size-fits-all tools.

\subsection{Mitigating Homogeneity of LLM Outputs}
Because the homogenization effect of ChatGPT seems to stem largely from the LLM generating similar responses to different users' queries in the same creative context, one obvious approach to reducing homogenization involves getting LLMs (or LLM-based CSTs) to output more diverse responses. Several strategies for achieving LLM output diversity might be implementable in the future.

At the CST user level, it may seem tempting to suggest that users of current LLMs can resist homogenization by implementing more sophisticated prompting strategies, such as deliberately designing prompts that incorporate some aspect of the user's unique perspective (to differentiate one's own prompt from the prompts one expects others to use in similar contexts). However, our analysis of prompting process (Section \ref{sec:ProcessImpact}) found that---although different ChatGPT users in our study did employ different prompting strategies---there were no reliable differences in homogenization to be found between users of different approaches to prompting. Additionally, LLM prompt design is difficult for novice users~\cite{WhyJohnnyCantPrompt}, and changes in prompt structure can dramatically alter both the relative and absolute performance of different LLMs on a single task~\cite{StateOfWhatArt}, suggesting that what makes for a good LLM prompt is likely opaque to users in general. Consequently, user adoption of more sophisticated prompting strategies may not yield reliable improvements in terms of group-level homogenization without additional interventions at the level of CST design or model development.

At the CST design level, injecting randomness into LLM prompts (for instance, by drawing stimuli at random from a large pool of potential stimuli---such as the Oblique Strategies deck---and then instructing the LLM to consider these random stimuli in responding to the user's input) might serve as a stopgap approach to increasing output diversity when different users are expected to input very similar prompts. However, due again to the complexity of prompting, thorough testing would need to be undertaken of any such prompt-level randomness injection strategy to ensure that it does in fact lead to increased output diversity.

Meanwhile, at the model level, strategies such as quality-diversity optimization (e.g., QDAIF~\cite{QDAIF}) and diverse decoding methods~\cite{DiverseDecodingIppolito,DiverseDecodingSee} may be used to improve the diversity of LLM responses. Compared to prompt-level randomness injection, these algorithmic strategies attempt to achieve output diversity more directly, and can thus be expected to yield more reliably diverse results in a wide variety of different creative contexts.

\subsection{Implications for LLM-Based CST Design}
Many of our findings follow naturally from the fact that an LLM will tend to produce similar outputs in response to similar inputs, even when those inputs come from different users. This property of LLMs is often desirable, especially when the goal is to build systems that respond reliably to queries in a particular domain. However, the fact that LLM outputs are strongly reflective of direct user inputs raises a key question for designers of LLM-based CSTs: when users are in the early stages of a creative process and their creative intent is still relatively undefined, how can an LLM-based tool avoid guiding them toward well-trodden ground?

A vaguely defined creative intent, expressed as (for instance) a short text string containing only a handful of words, will often tend---for level-of-detail reasons alone---to be similar to vaguely defined intents expressed by other users. For an LLM to directly transform a brief expression of intent into a substantially larger artifact (e.g., a passage of text), it will necessarily have to make a large number of creative decisions (e.g., individual word choices) on the user's behalf, and these decisions will be made by reference to the relative likelihood of different alternatives. Brief statements of creative intent therefore yield ``samey'' artifacts, which exhibit a phenomenon characterized elsewhere as the \emph{dearth of the author}~\cite{Dearth}. In other words, to get highly original outputs from an LLM, the user must supply the LLM with an input that contains enough detail to strongly differentiate it from other users' inputs; the LLM itself functions most directly as a means of refracting user-supplied inputs through the lens of statistical patterns learned from its training data.

In light of this limitation, we believe that it may be necessary to move toward a view of creativity support as \emph{intent elicitation}~\cite{IntentElicitation}: progressively drawing a more and more detailed, idiosyncratic specification of creative intent out of the CST user, with the goal of getting users to think \emph{more} rather than \emph{less} about each creative decision that they make along the way. By scaffolding the multi-turn discovery and expression of creative intent, we hope that LLM-based CSTs can eventually operate as something like a focusing lens for the imagination: a means of gradually clarifying and magnifying the spark of originality that first emerges in the user's head, without substituting the LLM's implicitly statistical and majoritarian process of creative decision-making for the user's own. Whether this design vision can be effectively realized remains to be seen, but we believe it to be an especially worthwhile goal for the next generation of AI-based CSTs to pursue.

\section{Limitations}
Our study was conducted in a lab context and consequently may not reflect some aspects of how people use LLMs for creativity support ``in the wild''. In order to keep study sessions to a manageable length, participants were only allowed a fixed and relatively short amount of time to respond to each ideation prompt; additionally, the ideation prompts in our study were supplied by us, rather than being formulated independently by CST users. Most organic divergent ideation scenarios are not subject to these constraints.

Because our study made use of a general-purpose LLM-based chatbot (ChatGPT) rather than an LLM-based CST specifically designed to support divergent ideation, it should be noted that alternative deployments of LLMs may be able to avoid homogenizing user ideas. \citeauthor{PadmakumarHe} have found evidence that feedback-tuned LLMs may induce creative homogenization where base (non-feedback-tuned) LLMs do not~\cite{PadmakumarHe}, and task-specific CST design has previously shown promise for reducing conformity effects in crowdsourced divergent ideation~\cite{DistributedAnalogicalIdeaGeneration}. These findings suggest that interventions at both the AI and UI layers may help to mitigate homogenization effects in future LLM-based CSTs.

\section{Conclusion}
We have presented evidence that LLM-based CSTs exert a stronger homogenization effect on human-in-the-loop divergent ideation processes than at least some plausible alternative CSTs. We have further clarified that the homogenization effect is group-level rather than individual-level, and that it may be partly attributable to low inferential distance between LLM outputs and apparently finished ideas. Coupled with evidence that ChatGPT users exhibit greater fluency, flexibility, and elaboration than users of an alternative CST, these results suggest that current general-purpose instruction-tuned LLMs (such as ChatGPT) are capable of functioning as useful CSTs by enabling the rapid enumeration of relatively obvious possibilities that users might otherwise fail, or take longer, to consider. However, these systems are not currently well-suited to helping users develop truly original ideas.

We believe that the style of homogenization analysis employed here is suitable for wider adoption as a technique for the evaluation of CSTs---including, but not limited to, AI-based CSTs---in the future. By providing both evidence for the existence of and proposed mechanisms of action for creative homogenization effects of AI-based CSTs, we also hope to stimulate future work on the adaptation of AI-based CSTs to mitigate these effects. Data-driven AI technologies clearly have a role to play in creative ideation processes, but interventions at both the CST design and model development levels may be required to realize the full potential of these technologies for creativity support.

\begin{acks}
This work was supported by Hackworth Grant GR102981, ``Investigating Homogenization of Imagination by Generative AI Models'', from the Markkula Center for Applied Ethics.
\end{acks}

\bibliographystyle{ACM-Reference-Format}


\appendix
\section{Validating Sentence Embeddings for Homogenization Analysis}\label{sec:ValidatingEmbeddings}
Our primary homogenization analysis uses a transformer-based sentence embedding model---\texttt{all-MiniLM-L6-v2}, one of the standard general-purpose sentence embedding models provided by the Python SentenceTransformers library~\cite{SentenceTransformers}---to evaluate the semantic similarity between participant ideas expressed as short strings of text. Our methodology here is similar to that employed in several other recent psychological studies of creativity. Broadly speaking, semantic similarity approaches to creativity research involve the use of some algorithm to produce a numeric score representing the similarity of a pair of creative artifacts; the originality of multiple different artifacts can then be determined and compared relative to a fixed reference point. The SemDis platform~\cite{SemDis}, a key example of this approach, automates large-scale determination of semantic similarity scores between ideas (expressed as short strings of text) and the creative ideation prompt in response to which these ideas were generated; similarity scores are determined by means of cosine similarity between aggregated word embeddings~\cite{GloVe}, a metric which has been found to agree well with human judgments of semantic similarity~\cite{SemDis,Dumas}.

Aggregated word embeddings are generally outperformed on semantic similarity tasks by more recent transformer-based sentence embeddings, which (unlike aggregated word embeddings) are able to take sentence structure into account. However, the rapid pace of progress in machine learning research means that transformer-based sentence embeddings have not yet been validated against human judgments of semantic similarity in the context of creativity research specifically. Therefore, in order to validate our use of \texttt{all-MiniLM-L6-v2}, we conducted a small experiment to determine whether this model agrees strongly with human judgments of semantic similarity on our participant ideas dataset.

Our experiment took the human-constructed idea categories produced by our flexibility analysis (Section \ref{sec:Flexibility}) as a source of ground truth for semantic similarity judgments and evaluated several candidate embedding models in terms of their agreement with the human coders' manual classification of ideas. First, for each category of ideas in our dataset, we produced a category embedding by averaging together the individual embeddings of the ideas belonging to this category. We then iterated over each idea in the dataset, sorted the category embeddings by their cosine similarity to the embedding of the idea being categorized, and assigned the idea to the $n$ categories represented by the $n$ most similar category embeddings (where $n$ = the number of categories human coders assigned to this idea). To avoid producing artificially high similarity scores between ideas and their actual human-assigned categories across the board, we also excluded each idea's own embedding from the average category embedding when testing similarity to the idea's actual human-assigned categories.

We then compared the model-assigned categories for each idea to the actual categories human coders assigned to this idea, and noted the percentage of overlap between these category sets. Finally, we repeated this process for several different embedding models---as well as a pessimistic baseline ``model'' that assigned each idea to $n$ categories at random---and computed the human-agreement percentage of each model on participant ideas generated in response to each of our four creativity tasks.

Results are reported in Table \ref{tab:embeddings}. Notably, our chosen sentence embedding model (\texttt{all-MiniLM-L6-v2}) agrees with human idea categorizations more than half the time across all four creativity tasks; it therefore substantially outperforms both GloVe 840B (an aggregate word embedding model previously assessed as state-of-the-art for creativity research~\cite{SemDis,Dumas}) and the random baseline (which GloVe itself beats by more than an order of magnitude). It also consistently outperforms \texttt{all-mpnet-base-v2}, the theoretically best overall general-purpose SentenceTransformers model, by a small margin. 

\begin{table}[]
    \centering
    \begin{tabular}{c|cccc|c}
    \textbf{Model} & \textbf{IC\_A} & \textbf{IC\_B} & \textbf{PI\_A} & \textbf{PI\_B} & \textbf{Average} \\
    \hline
    all-MiniLM-L6-v2 & 60.94\% & 58.61\% & 51.48\% & 64.63\% & 58.92\% \\
    all-mpnet-base-v2 & 58.91\% & 57.08\% & 51.30\% & 59.78\% & 56.77\% \\
    GloVe 840B & 46.59\% & 47.25\% & 41.16\% & 40.02\% & 43.76\% \\
    Random & 4.31\% & 2.37\% & 4.14\% & 2.94\% & 3.44\%
    \end{tabular}
    \caption{Percentage agreement of several different embedding models with human idea categorization judgments. Columns IC\_A, IC\_B, PI\_A, and PI\_B report performance on ideas generated in response to specific ideation tasks (IC = ``Improbable Consequences'', PI = ``Product Improvement'').}
    \label{tab:embeddings}
\end{table}

We did not directly evaluate OpenAI's sentence embedding models~\cite{OpenAIEmbeddings}, despite the use of an unspecified OpenAI embedding model in another study of creative homogenization~\cite{DoshiHauser}, because OpenAI models (unlike the SentenceTransformers models) are proprietary: they cost money to use, cannot be run locally, and are not open source. This limits the replicability of research results derived from OpenAI models and therefore the suitability of these models for research pipelines. However, due to the prior use of one of these models in a similar research context, a comparison of some sort is nevertheless merited. For this, we consult the Massive Text Embedding Benchmark~\cite{MTEB}, which shows that the most powerful OpenAI embeddings model (\texttt{text-embedding-ada-002}) is not clearly much better or worse than a variety of SentenceTransformers models (including \texttt{all-MiniLM-L6-v2}) across a range of tasks.

Sentence embedding models remain imperfect arbiters of semantic similarity. In our categorization experiment, even the best-performing embeddings model achieved only 59\% agreement with human coders on average. There also exists some evidence that cosine similarity between sentence embeddings is more strongly influenced by overlap in the set of nouns than by other similarities~\cite{SentenceTransformerBias}, suggesting that these models do not take all of the nuances of sentence meaning into account when computing similarity scores. However, the agreement between models like \texttt{all-MiniLM-L6-v2} and human judgments of semantic similarity strike us as high enough to justify the use of these sentence embedding models for homogenization analysis in creativity research, in particular for the increased scale of analysis that these models enable.

\end{document}